\documentclass[11pt,twoside]{article}

\usepackage{asp2006}
\usepackage{epsf}
\usepackage{psfig}
\usepackage{lscape}
\usepackage{graphicx}

\def \etal {et~al.~}

\begin{document}
\title{The Impact of Feedback on Disk Galaxy Scaling
  Relations} 
\author{Aaron A. Dutton$^{1}$ and Frank C. van den Bosch$^{2}$}
\affil{
  $^{1}$ UCO/Lick Observatory, University of California,  Santa Cruz, CA 95060 \\
  $^{2}$ Department of Physics and Astronomy, University of Utah, 115 South 1400 East, Salt Lake City, UT 84112-0830\\
}

\begin{abstract} We use a disk formation model to study the effects of
  galactic outflows (a.k.a. feedback) on the rotation velocity -
  stellar mass - disk size, gas fraction - stellar mass, and gas
  phase metallicity - stellar mass scaling relations of disk galaxies.
  We show that models without outflows are unable to explain these
  scaling relations, having both the wrong slopes and
  normalization. The problem can be traced to the model galaxies
  having too many baryons. Models with outflows can solve this
  ``over-cooling'' problem by removing gas before it has time to turn
  into stars. Models with both momentum and energy driven winds can
  reproduce the observed scaling relations. However, these models
  predict different slopes which, with better observations, may be
  used to discriminate between these models.
\end{abstract}

\section{Introduction}  

Galactic outflows are widely observed in galaxies that are undergoing,
or have recently undergone, intense star formation: e.g. Nearby
starburst and IR bright galaxies (Martin 2005); Post starburst
galaxies at redshift $z\simeq 0.6$ (Tremonti \etal 2007); Normal Star
forming galaxies at redshift $z=1.4$ (Weiner \etal 2009) and Lyman
Break Galaxies at redshifts $z\simeq 3$ (Shapley \etal 2003).
However, whether or not galactic outflows play an important role in
determining the properties of galaxies has yet to be determined.

A clue that outflows might play an important role in galaxy formation
comes from fact that galaxy formation is inefficient. The galaxy
formation efficiency, $\epsilon_{GF}$, defined as the ratio between
the galaxy mass (in stars and cold gas) to the total available baryons
available to that galaxy (the cosmic baryon fraction times total virial
mass of the halo) peaks at $\simeq 33\%$. This has been determined by
galaxy-galaxy weak lensing studies (Hoekstra \etal 2005; Mandelbaum
\etal 2006), which are independent of $\Lambda$CDM, and galaxy-halo number
abundance matching (e.g. Conroy \& Wechsler 2009), which assumes
the $\Lambda$CDM halo mass function as a prior.

A low peak galaxy formation is a problem because cooling is expected
to be efficient in typical galaxy mass haloes (with virial velocities
ranging from $V_{\rm vir} \simeq 60$ to $\simeq 150\, \rm km/s$).  At
low masses (below $V_{\rm vir} \simeq 30 \,\rm km/s$) cooling is
suppressed by UV photo heating, while at high masses (and high virial
temperatures) cooling is inefficient due to the physics of radiative
cooling.  Thus another mechanism is needed to suppress galaxy
formation, in the halo mass regime one would expect it to be highly
efficient. Galactic outflows driven by supernova (SN) or young massive
stars are the prime candidate, having been successfully invoked in
semi-analytic galaxy formation models to explain the shallow faint end
of the galaxy luminosity function (e.g. Benson \etal 2003).

\subsection{Simple feedback models}
The simplest, physically motivated, feedback models can be described
by 2 parameters: the mass loading factor, $\eta$, defined as the ratio
between the mass outflow rate, and the star formation rate; and the
wind velocity, $V_{\rm wind}$. These two parameters are related by the
mechanism that drives the wind, and the relation between the wind
velocity and the escape velocity, $V_{\rm esc}$.  Feedback models can be
divided into 3 broad categories:

\begin{itemize}
\item {\bf Constant Velocity Wind}: Assumes $V_{\rm wind}=const.$,
  which implies $\eta=const$. A popular example is that implemented by
  Springel \& Hernquist (2003), which assumes $V_{\rm wind}=484 \rm
  km/s$ and $\eta=2$. This corresponds to 25\% of the SN energy being
  transferred to the wind (i.e. $\epsilon_{\rm FB}=0.25$).
\item {\bf Momentum Driven Wind}: Assumes $V_{\rm wind}=3\sigma\simeq
  V_{\rm esc}$, where $\sigma$ is the velocity dispersion of the
  galaxy. Momentum conservation implies $\eta = (300/V_{\rm wind})$ (this
  assumes 100\% momentum conservation) (Murray Quataert \& Thompson
  2005).
\item {\bf Energy Driven Wind}: Assumes $V_{\rm wind}=V_{\rm esc}$,
  energy conservation implies $\eta=\epsilon_{\rm FB}\, 10 (300/V_{\rm
    wind})^2$, where $\epsilon_{\rm FB}$ is the fraction of SN energy
  that ends up in the outflow (e.g. van den Bosch 2001)
\end{itemize}  

Finlator \& Dav{\'e} (2008) showed that models with the momentum
driven wind provide a better match to the stellar mass - gas phase
metallicity relation at $z\simeq 2$ than models with a constant
velocity energy driven wind, or models without galaxy winds. However,
it is not clear that this is a convincing argument against energy
driven winds because Finlator \& Dav{\'e} (2008) did not consider an
energy driven wind with the same assumption that they made for the
momentum driven wind i.e.  $V_{\rm wind} \simeq V_{\rm esc}$.

Here we use a semi-analytic disk galaxy formation model to discuss the
observational signatures of different feedback models on the scaling
relations of disk galaxies. We address the following questions: 1) Can
models without outflows explain these relations?  2) Can models with
outflow explain these relations? and 3) Can the scaling relations be
used to discriminate between different wind models?

\section{The Disk Galaxy Formation Model}  
\label{sec:models}
Here we give a brief overview of the disk galaxy evolution model used
in this proceedings. This model is described in detail in Dutton \&
van den Bosch (2009).  The key difference with almost all disk
evolution models is that in this model the inflow (due to gas
cooling), outflow (due to SN driven winds), star formation rates, and
metallicity, are computed {\it as a function of galacto centric
  radius}, rather than being treated as global parameters.  The main
assumptions that characterize the framework of these models are the
following:
\begin{enumerate}
\item {\bf Mass Accretion History}: Dark matter haloes around disk
  galaxies grow by the smooth accretion of mass which we model with
  the Wechsler \etal (2002) mass accretion history (MAH). The shape of
  this MAH is specified by the concentration of the halo at redshift
  zero;
\item {\bf Halo Structure}: The structure of the halo is given by the NFW
  profile (Navarro, Frenk, \& White 1997), which is specified by two
  parameters: the mass and concentration. The evolution of the
  concentration parameter is given by the Bullock \etal (2001) model
  with parameters for a WMAP 5th year cosmology 
  (Macci\`o \etal 2008);
\item {\bf Angular Momentum}: Gas that enters the halo is
  shock heated to the virial temperature, and acquires the same
  distribution of specific angular momentum as the dark matter.  We
  use the angular momentum distributions of the halo as parametrized
  by Sharma \& Steinmetz (2005);
\item {\bf Gas Cooling}: Gas cools radiatively, conserving its specific
  angular momentum, and forms a disk in centrifugal equilibrium;
\item {\bf Star Formation}: Star formation occurs according to a
  Schmidt type law on the dense molecular gas, which is computed
  following Blitz \& Rosolowsky (2006);
\item {\bf Supernova Feedback}: Supernova feedback re-heats some of the
  cold gas, ejecting it from the disk and halo;
\item {\bf Metal Enrichment}: Stars eject metals into the inter stellar
  medium, enriching the cold gas.
\item {\bf Stellar Populations}: Bruzual \& Charlot (2003) stellar
  population synthesis models are convolved with the star formation
  histories and metallicities to derive luminosities and surface
  brightness profiles.
\end{enumerate}


\begin{figure}[!ht]
\centerline{
\includegraphics[scale=0.49]{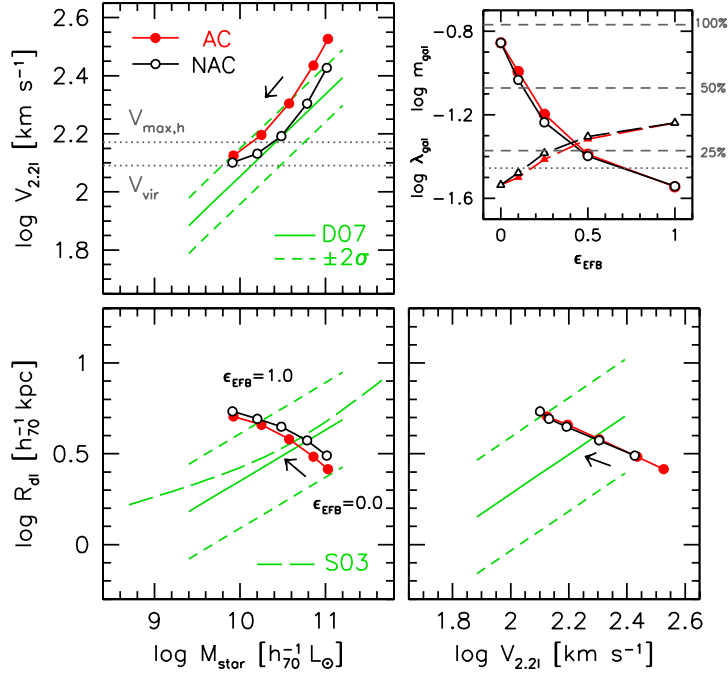}}
\caption{{\footnotesize \bf Dependence of Velocity, Stellar Mass and
    Disk Size on Feedback:} Effect of feedback efficiency,
  $\epsilon_{\rm FB}$, for the energy driven wind, on the position of
  a galaxy with $M_{\rm vir} = 6.3 \times 10^{11} h^{-1} M_{\odot}$ in
  the $VMR$ planes.  The arrows indicate the direction of increasing
  $\epsilon_{\rm FB}$.  Models with adiabatic contraction are shown
  with solid red symbols, models without adiabatic contraction are
  shown as black open symbols.  The solid and dashed green lines in
  show the mean and 2$\sigma$ scatter of the observed relations from
  Dutton \etal (2007, D07), assuming a Chabrier IMF. The long dashed
  green line shows the observed half-light radius stellar mass
  relation from Shen \etal (2003, S03).  The panel in the top right
  shows the effect of feedback on the galaxy mass fraction, $m_{\rm
    gal}$ (circles), and galaxy spin parameter, $\lambda_{\rm gal}$
  (triangles).  The dashed horizontal lines show galaxy formation
  efficiencies of 100, 50, and 25 percent, the dotted horizontal line
  shows the spin parameter of the halo.  As the feedback efficiency is
  increased the galaxy mass fraction ($m_{\rm gal}$) decreases, the
  galaxy spin parameter ($\lambda_{\rm gal}$) increases.  This results
  in the rotation velocity decreasing, the stellar mass decreasing,
  and the size of the stellar disk increasing.}
\end{figure}

\section{Results}

\subsection{Impact of Feedback on Velocity, Stellar Mass and
  Disk Size}
Fig.~1 shows the impact of feedback on the rotation velocity, stellar
mass, and disk size of a galaxy that forms in a halo with virial mass
$M_{\rm vir} = 6.3 \times 10^{11} h^{-1} M_{\odot}$, and which has the
median halo concentration and angular momentum parameters for haloes
of this mass.  The green lines show the observed scaling relations
from (Dutton \etal 2007 and Shen \etal 2003). The circles show models
with feedback efficiency varying from $\epsilon_{\rm FB}=0$ to 1. The
model without feedback results in a galaxy that is too small and which
rotates too fast.  The upper right panel shows the galaxy mass
fraction $m_{\rm gal} = M_{\rm gal}/M_{\rm vir}$, and galaxy spin
parameter $\lambda_{\rm gal} = (j_{\rm gal}/m_{\rm gal}) \lambda$,
where $\lambda$ is the spin parameter of the halo and $j_{\rm gal} =
J_{\rm gal}/J_{\rm vir}$ is the angular momentum fraction of the
galaxy, versus the feedback efficiency. This shows that the model
galaxy without feedback has acquired 85\% of the available baryons and
80\% of the available angular momentum. The mass and angular momentum
fractions are less than unity because cooling is not 100\%
efficient. The angular momentum fraction is less than the galaxy mass
fraction because cooling occurs from the inside-out.

As the feedback efficiency is increased the galaxy stellar mass
decreases, disk size increases and the rotation velocity decreases.
These changes are primarily driven by the decrease in the galaxy mass
fraction, $m_{\rm gal}$, and secondarily by the increase in the galaxy
spin parameter, $\lambda_{\rm gal}$ (upper right panel). The increase
in galaxy spin parameter is the result of preferential loss of low
angular momentum material, which helps to explain the origin of
exponential galaxy disks, which are otherwise not naturally produced
in a CDM cosmologies (Dutton 2009).

The upper left panel shows that models with adiabatic contraction
(Blumenthal \etal 1986) (red points and lines) rotate too fast for all
feedback efficiencies. For a model without adiabatic contraction (open
circles and black lines) the zero point of the VM relation is
reproduced for feedback efficiencies of $\epsilon_{\rm FB} \simeq
0.1-0.5$. In order for our models to produce realistic rotation
velocities, in the models that follow we will assume the halo does not
contract in response to galaxy formation.

\begin{figure}[!ht]
\centerline{
\includegraphics[scale=0.55]{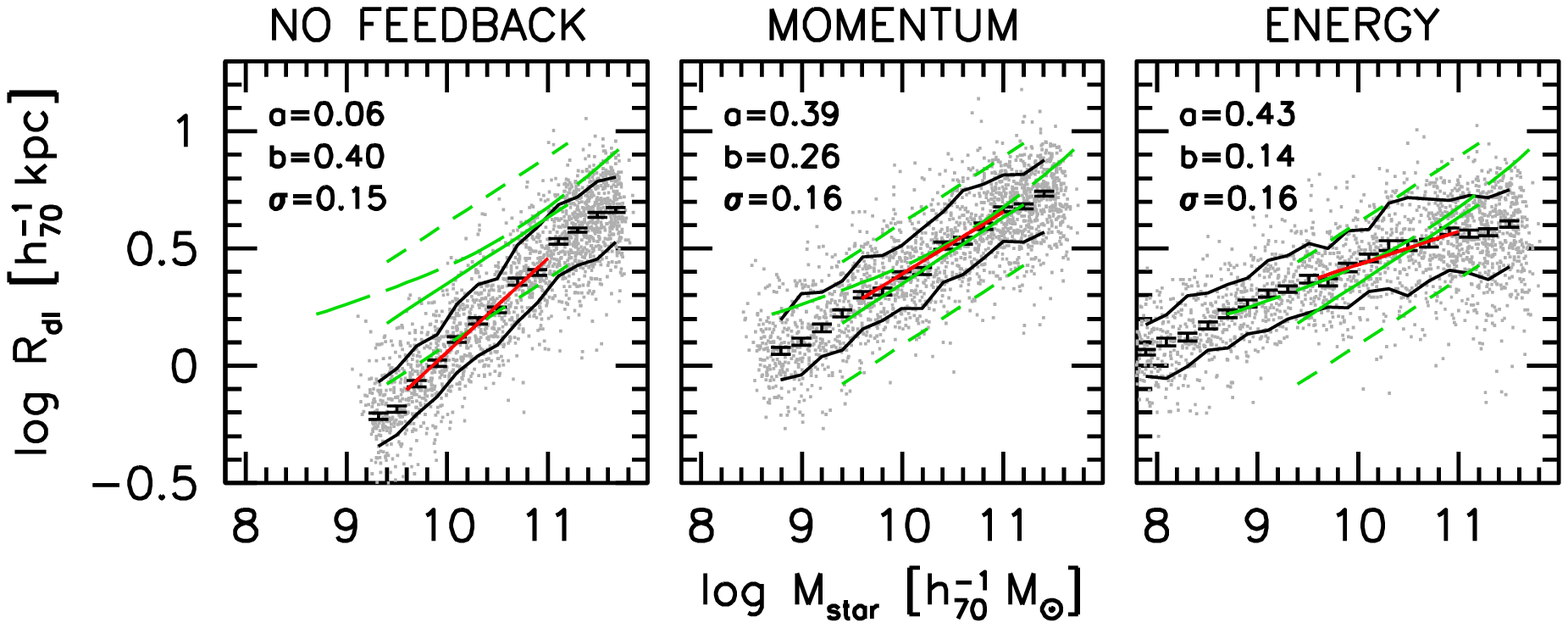}}
\centerline{
\includegraphics[scale=0.55]{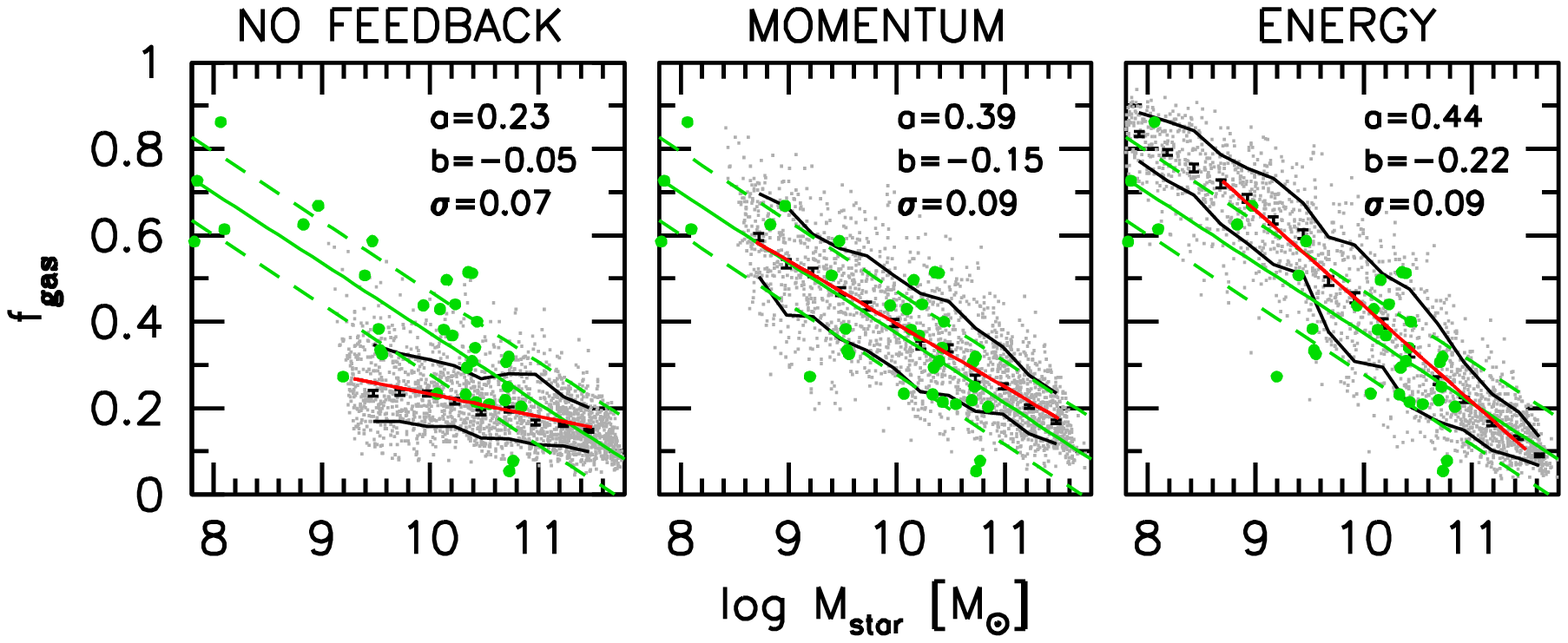}
}
\centerline{
\includegraphics[scale=0.55]{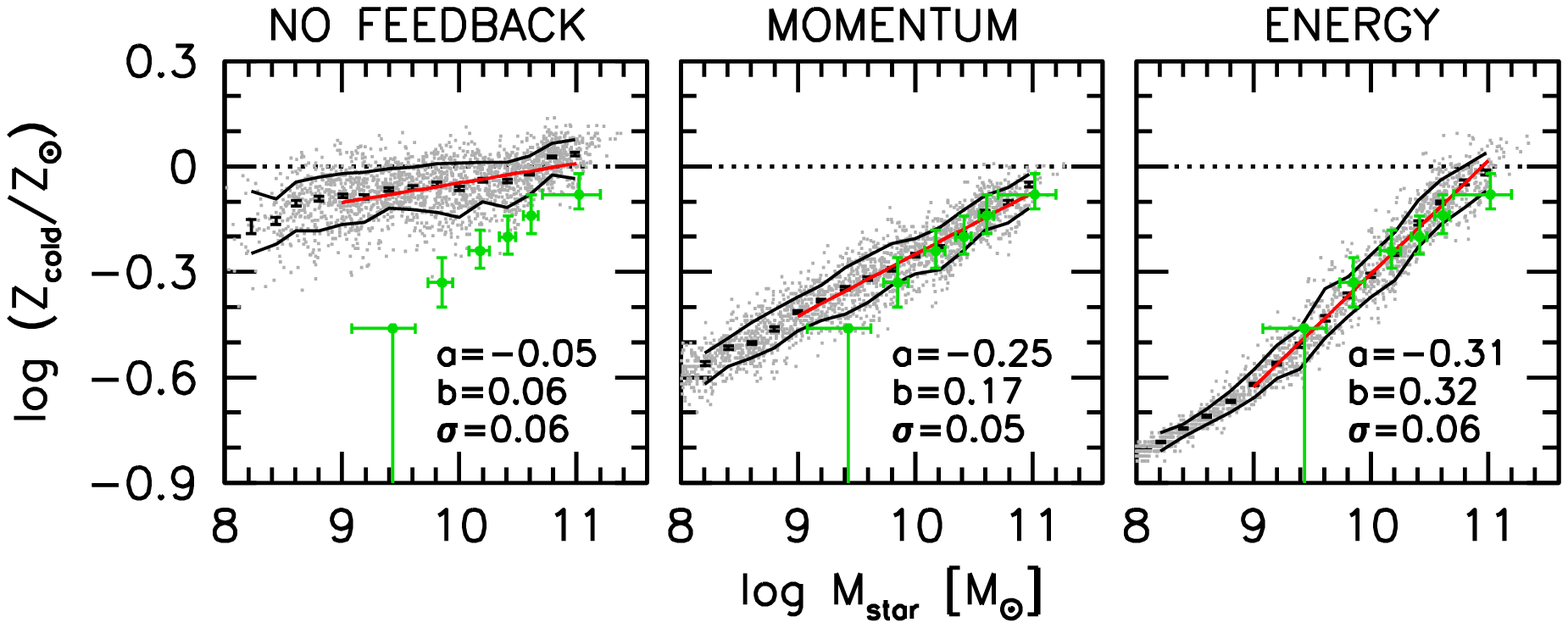}
}
\caption{\footnotesize {\bf Dependence of disk size, gas fraction and
    gas metallicity on feedback.} {\it Upper panels:} disk size -
  stellar mass; {\it Middle panels:} gas fraction - stellar mass, {\it
    Lower panels:} gas phase metallicity - stellar mass. The observed
  relations are given by green lines, points and symbols. The model
  galaxies are given by grey points, with the black lines showing the
  14th and 86th percentiles in stellar mass bins.  For the size-mass
  relation and gas fraction mass scaling relations the data (Dutton
  \etal 2007; Shen \etal 2003; Garnett 2002) and models are for
  redshift $z=0$. For the metallicity-mass relation the data (Erb
  \etal 2006) and models are for redshift $z=2.26$. The sizes, gas
  fractions and metallicities are coupled, and yield different slopes
  for different feedback models}
\label{fig:fig1}
\end{figure}

\subsection{Impact of feedback on disk sizes, gas fractions and metallicity}
Here we discuss the impact of feedback on the scaling relations
between disk size, gas fractions and gas phase metallicity with
stellar mass. We discuss three feedback models: 1) no feedback; 2)
momentum driven feedback; 3) energy driven feedback with
$\epsilon_{\rm FB}=0.25$. For each model we generate a Monte Carlo
sample of galaxies, with halo masses logarithmically sampled from
$M_{\rm vir} =10^{10} - 10^{13} \,h^{-1} M_{\odot}$, log-normal scatter
in halo spin parameter $\lambda$, halo concentration, $c$, and angular
momentum distribution shape, $\alpha$.

{\bf Disk Sizes:} The upper panels of Fig.~2 show the disk size-
stellar mass relation for our three models. As expected from Fig.~1,
the model without feedback produces a size-mass relation with the
wrong zero point, but also with the wrong slope.  Models with feedback
reproduce the zero point of the size-mass relation, but they predict
different slopes: 0.26 for the momentum driven wind and 0.14 for the
energy driven wind. The energy driven wind predicts a shallower slope
because it is more efficient at removing gas from lower mass haloes,
which (see Fig.~1) moves galaxies to lower masses and larger sizes.
Observationally the correct slope is not clear, with values of 0.24
(Pizagno \etal 2005) and 0.28 (Dutton \etal 2007) and 0.14 (at low
masses) to 0.39 (at high masses) from Shen \etal (2003) being
reported. Thus a more accurate observational determination of the
size-stellar mass relation would provide useful constraints to these
models.

{\bf Gas Fractions:} It has emerged in the last few years (Springel \&
Hernquist 2005; Hopkins \etal 2009) that the gas fraction of galaxies
plays an important role in determining the morphology of galaxies
after mergers. In particular galaxies with high gas fractions can
re-form their disks after major and intermediate mass mergers. This
removes a potential problem for the formation of bulgeless and low
bulge fraction galaxies in $\Lambda$CDM, where intermediate and major
mergers occur in the lifetime of essentially all dark matter haloes.

The middle panels of Fig.~2 show the gas fraction vs. stellar mass
relation. The green points show observations from Garnett (2002), with
a fit to the mean and scatter of this data shown by the solid and
dashed lines.  The model without feedback (left) produces galaxies
that are too gas poor, especially for lower mass galaxies. This
problem is the result of the disks being too small, and hence too high
surface density, which results in more efficient star formation. The
models with feedback both reproduce the observed relation, with the
energy driven wind predicting a higher zero point.

{\bf Mass Metallicity:} Finlator \& Dav{\'e} (2008) used the mass
metallicity relation at redshift $z \simeq 2$ to argue in favor of
momentum driven winds over energy driven winds (of constant
velocity). The lower panels of Fig.~2 show the stellar mass - gas
metallicity relation at $z=2.26$. We confirm the result of Finlator \&
Dav{\'e} (2008) that models without feedback do not reproduce the
mass-metallicity relation, and that models with momentum driven winds
provide a good match to the observations. However, we also show that
models with energy driven winds provide a equally good match to the
data.  The energy and momentum driven winds do predict different
slopes: 0.17 for momentum and 0.32 for energy, and thus more accurate
observations, and especially to lower stellar masses, may be able to
distinguish between these two models.

\section{Summary}
We have used a semi-analytic disk galaxy formation model to
investigate the effects of galaxy outflows on the scaling relations of
disk galaxies. We find that 

{\bf 1)} None of the scaling relations can be reproduced in models
without outflows: model galaxies rotate too fast, are too small, are
too gas poor and are too metal rich.  These problems are driven by the
high baryonic mass fractions of these galaxies.

{\bf 2)} Models with outflows can solve this problem by removing gas
from galaxies before it has had time to turn into stars.

{\bf 3)} Models with momentum and energy driven winds provide
acceptable fits to the observed disk size-stellar mass, gas fraction
stellar mass, and gas metallicity - stellar mass relations. However,
these models predict different slopes (due to the different scaling
between mass loading factor and wind velocity). Thus more accurate
observations will be able to discriminate between these models.




\end{document}